\documentclass[a4paper,aps,prd,floatfix,reprint,twoside,twocolumn,preprintnumbers,showkeys,showpacs,final]{revtex4-2}
\usepackage[utf8]{inputenc}
\usepackage{amsmath, amssymb}
\usepackage{siunitx}
\usepackage{graphicx}
\usepackage[caption=false]{subfig}
\usepackage[section]{placeins}
\usepackage{xcolor}
\usepackage{hyperref} 

\usepackage{environ} \usepackage{tikz}
\usetikzlibrary{arrows,decorations.markings}
\usetikzlibrary[arrows.meta,bending]
\usetikzlibrary{positioning}

\makeatletter \newsavebox{\measure@tikzpicture}
\NewEnviron{scaletikzpicturetowidth}[1]{
  \def\tikz@width{#1}
  \begin{lrbox}{\measure@tikzpicture}
    \BODY
  \end{lrbox}
  \pgfmathparse{#1/\wd\measure@tikzpicture}
  
  \BODY } \makeatother

\newcommand{\vect}[1]{\ensuremath{\mathbf{#1}}}
\DeclareMathOperator{\Tr}{Tr} 

\begin{document}
\preprint{ADP-22-23/T1194}
\title{Impact of Dynamical Fermions on the Centre Vortex Gluon Propagator} \author{James C.
  Biddle} \author{Waseem Kamleh} \author{Derek B. Leinweber} \affiliation{Centre
  for the Subatomic Structure of Matter, Department of Physics, The University
  of Adelaide, SA 5005, Australia}

\begin{abstract}
The impact of $SU(3)$ centre vortices on the Landau-gauge gluon propagator is calculated in the presence of dynamical fermions and compared to the pure Yang-Mills case. The presence of dynamical fermions is found to alter the behaviour of the centre vortex propagator when compared to the established pure-gauge result. The gluon spectral representation is also explored from the centre vortex perspective, where centre vortices are shown to exhibit clear signs of positivity violation, which is an indicator of confinement. Vortex removal subsequently restores positivity, demonstrating the crucial role centre vortices play in the confinement of gluons.
\end{abstract}

\maketitle
\section{Introduction}\label{sec:Intro}
The first step in any lattice calculation is to simulate the ground state QCD
fields that permeate the vacuum. These fields give rise to the fundamental QCD
features of confinement and dynamical chiral symmetry breaking resulting in
dynamical generation of mass. Naturally, there has been substantial effort from
the community to deduce what aspect of these fields gives rise to these
features. Techniques have been developed to identify topological defects such as
Abelian monopoles~\cite{tHooft:1981bkw,Mandelstam:1974pi},
instantons~\cite{Belavin:1975fg,Witten:1978bc,Callan:1977gz,Aharonov:1978jd} and
centre
vortices~\cite{tHooft:1977nqb,tHooft:1979rtg,Cornwall:1979hz,Nielsen:1979xu}
within these ground state fields that present possible answers to this question.
Centre vortices are of particular interest as they offer access to the most
fundamental mechanism that could underpin these phenomena.

The centre vortex picture of QCD has experienced a renewed interest in recent
years due to a host of lattice results reinforcing their importance in an
understanding of the non-perturbative properties of QCD. The majority of these
studies have been performed in the context of pure-gauge QCD, where the effects
of fermion loops are omitted from the lattice Monte-Carlo procedure. These
results have shown that vortex removal results in a loss of dynamical mass
generation~\cite{Trewartha:2015nna,OMalley:2011aa,Trewartha:2017ive}, loss of
string tension~\cite{Langfeld:2003ev,Bowman:2010zr} and the suppression of the
infra-red landau gauge gluon propagator~\cite{Biddle:2018dtc,Bowman:2010zr}.
Vortices alone have been capable of reproducing the qualitative picture of
non-perturbative QCD through reproduction of the linear static quark
potential~\cite{Langfeld:2003ev,OCais:2008kqh,Trewartha:2015ida}, an infrared
dominated gluon propagator~\cite{Biddle:2018dtc}, and the re-introduction of
dynamical mass generation in the low-lying hadron
spectrum~\cite{Trewartha:2017ive}.

Despite the success of these results, there has been a persistent quantitative
discrepancy between centre-vortex results and those from the original gauge
fields on which the vortices are identified. This manifests in a variety of
ways, most notably in a lower string tension obtained from the static quark
potential~\cite{Langfeld:2003ev,OCais:2008kqh,Trewartha:2015ida}, lower hadron
masses in the low-lying hadron spectrum~\cite{Trewartha:2017ive}, and in residual
infrared strength being retained by the vortex-removed gluon
propagator~\cite{Biddle:2018dtc}. The origin of these discrepancies is as-yet
unknown, but they suggest a common property present in existing vortex studies.

Recently, work has been done to examine the effect of dynamical fermions on the
structure of centre vortices and the corresponding impact on the static quark
potential~\cite{Biddle:2022zgw,Virgili:2022ybm}. These results showed for the first time a
quantitative agreement between vortex-only results and those obtained from
unmodified gauge fields. These results motivate further exploration of the
relationship between centre vortices and dynamical fermions. Here we continue
this line of investigation by calculating the Landau gauge gluon propagator on
vortex-modified configurations in the presence of dynamical fermions.

We will also examine the gluon spectral density by calculating the Euclidean
correlator to determine the presence or absence of positivity violation, as
positivity violation serves as an indicator of gluon
confinement~\cite{Bowman:2007du}. It is well understood that positivity
violation in the gluon and quark propagators is a necessary condition for
light-quark confinement~\cite{Roberts:2007jh}. As such, positivity violation
arising from centre vortices serves as a strong indication that the centre
vortex mechanism underpins the confinement of physical particles.

This paper is structured as follows. Section~\ref{sec:Vortex-Identification}
outlines the vortex identification procedure on the lattice and describes the
ensembles used in this work. Section~\ref{sec:Gluon-Propagator} introduces the
gluon propagator and presents the results from our centre-vortex studies.
Section ~\ref{sec:Positivity-Violation} discusses the notion of positivity
violation and presents results based on an examination of the Euclidean
correlator. Section~\ref{sec:Conclusion} will summarise the findings of this
paper.

\section{Vortex Identification}\label{sec:Vortex-Identification}
The procedure of centre vortex identification on the lattice is now
well-established~\cite{Biddle:2018dtc,Biddle:2019gke}. To identify the `thin'
lattice vortices that correspond to the centre of physical `thick'
vortices~\cite{Engelhardt:1999xw,Bertle:2000ap}, it is necessary to bring all configurations in an ensemble into maximal centre gauge (MCG). This is done by applying a gauge transformation $\Omega(x)$ such that the functional
\begin{equation}
  \label{eq:1}
  R = \frac{1}{V\, N_\text{dim}\, n_c^2} \sum_{x, \mu}\left| \Tr U^{\Omega}_{\mu}(x) \right|^2\, ,
\end{equation}
is maximised. This serves to bring every gauge link $U_{\mu}(x)$ as close as possible to a centre element of $SU(3)$, where the group centre is comprised of the three elements proportional to the identity
\begin{equation}
  \mathbb{Z}_3 =  \exp\left(m\frac{2\pi i}{3} \right)I, ~ m\in\lbrace-1, 0, +1\rbrace .
\end{equation}
A more detailed description of the algorithm used to perform this updating procedure can be found in Ref.~\cite{Biddle:2022zgw}.

Once fixed to MCG, a new ensemble is constructed by projecting each link onto its nearest centre element. This centre-projected configuration is known as the vortex-only configuration, $Z_{\mu}(x)$. From this construction, we also define the vortex-removed ensemble as $R_{\mu}(x)=Z_{\mu}^{\dagger}(x)\,U_{\mu}(x)$. The result of this procedure is three ensembles:
\begin{enumerate}
\item Original, untouched (UT) fields, $U_{\mu}(x)$,
\item Vortex-only (VO) fields, $Z_{\mu}(x)$,
\item Vortex-removed (VR) fields, $R_{\mu}(x)$,
\end{enumerate}
We refer to the latter two collectively as the vortex-modified ensembles. It is these three ensembles that we study to determine the impact of centre vortices.

For this work, we make use of three original (UT) ensembles of 200
$32^3\times 64$ lattice gauge fields. Two of these are $(2 + 1)$ flavour
dynamical ensembles from the PACS-CS collaboration~\cite{Aoki:2008sm}. The
heaviest pion mass of $701~\si{MeV}$ and lightest pion mass of $156~\si{MeV}$
are chosen to provide the greatest range of masses to best see the impact of
centre vortices as the physical point is approached. The third ensemble is pure
Yang-Mills, and provides a reference point to previous studies. This pure-gauge
ensemble has been tuned to have a similar lattice spacing as the dynamical
ensembles so that finite volume effects should be similar across all ensembles
used in this work. A summary of the ensemble parameters is provided in
Table~\ref{tab:LatticeParams}.
\begin{table}[tb]
  \caption{\label{tab:LatticeParams}A summary of the lattice ensembles used in
    this work~\cite{Aoki:2008sm}.}
  \begin{ruledtabular}
    \begin{tabular}{ccccc}
      Type & $a\, (\si{fm})$ & $\beta$ & $\kappa_{\rm u,d}$ & $m_{\pi}\, (\si{MeV})$ \\
      \hline\\
      Pure gauge & 0.100 & 2.58 & - & - \\
      Dynamical & 0.102 & 1.9 & 0.13700 & 701\\
      Dynamical & 0.093 & 1.9 & 0.13781 & 156
    \end{tabular}
  \end{ruledtabular}
\end{table}
\section{Gluon Propagator}\label{sec:Gluon-Propagator}
\subsection{Definition}\label{subsec:GlupropDefinition}
In the continuum, the momentum-space Landau gauge gluon propagator is of the form
\begin{equation}
  \label{eq:Gprop-Cont}
  D^{ab}_{\mu\nu}(q) = \left ( \delta_{\mu\nu} - \frac{q_\mu q_\nu}{q^2} \right )\,\delta^{ab}\,D(q^2) \, ,
\end{equation}
where where $D(q^2)$ is the scalar gluon propagator. On the lattice, the scalar propagator for $p^{2}\ne 0$ is calculated by considering~\cite{Biddle:2018dtc}
\begin{equation}
  \label{eq:Scalar-Propagator}
  D(p^2) = \frac{2}{3\,(n_c^2-1)\,V}\big\langle {\rm Tr}\, A_\mu(p)\,A_\mu(-p) \big\rangle \,.
\end{equation}
where $n_c=3$ is the number of colours, $V$ is the lattice volume and $A_\mu(p)$ is calculated via the discrete Fourier transform of the
midpoint definition of the gauge potential~\cite{Alles:1996ka},
\begin{multline}
A_\mu(x+\hat{\mu}/2)=\frac{1}{2i}\, \left (U_\mu(x)-U^{\dagger}_\mu(x)\right ) \\
-\frac{1}{6 i}\, {\rm Tr}\left (U_\mu(x)-U^{\dagger}_\mu(x) \right ) + {\cal O}(a^2)\,.
\end{multline}
As the gauge fields used in this analysis are generated using the
$\mathcal{O}(a^2)$-improved Iwasaki action~\cite{Iwasaki:1983iya}, the tree-level
behaviour of the gluon propagator is improved by making the substitution~\cite{Symanzik:1983dc,Weisz:1983bn,Weisz:1982zw}
\begin{equation}
  \label{eq:Momentum-Substitution}
  p_\mu \rightarrow q_\mu = \frac{2}{a}\sqrt{\sin^2\left(\frac{p_\mu a}{2}\right)+\frac{1}{3}\sin^4\left(\frac{p_\mu a}{2}\right)}\, ,
\end{equation}
where $p_\mu$ are the usual lattice momentum variables
\begin{equation}
  \label{eq:Momentum-Variables}
p_\mu = \frac{2\pi\,n_\mu}{aN_\mu}, ~~n_\mu \in \left(-\frac{N_\mu}{2},\frac{N_\mu}{2}\right]\, ,
\end{equation}
and $N_\mu$ is the lattice extent in the $\mu$ direction. The tree-level continuum scalar propagator is then given by $D(q^2)=\frac{1}{q^2}$. This choice of momentum variables reduces the sensitivity of the gluon propagator to finite lattice spacing effects at large momenta~\cite{Bonnet:2001uh}.

The perturbative scalar propagator is defined as $D(q^2)=\frac{Z(q^2)}{q^2}$. For the remainder of this section we will focus on the renormalisation function $Z(q^2) = q^2\,D(q^2)$. We then renormalise $Z(q^2)$ in the momentum space subtraction (MOM) scheme~\cite{Leinweber:1998uu,Bowman:2004jm} on the untouched configurations by enforcing the condition that $Z^{\rm UT}(\mu^2) = 1$ at the largest available momentum on all ensembles, $\mu=5.5~\si{GeV}$. This is performed via determination of a constant $Z_3^{\rm UT}$ satisfying
\begin{equation}
  \label{eq:MOMRenorm}
  \frac{Z^{\rm UT}_{\rm bare}(\mu^{2})}{Z_3^{\rm UT}} = Z^{\rm UT}(\mu^{2})=1\,.
\end{equation}

Renormalising the vortex-modified results requires more careful consideration, as there is no \textit{a priori} method by which it should be performed. Specifically, the problem arises from the absence of a perturbative expectation for the vortex-only propagator. The vortex-removed results are expected to encapsulate the high-momentum behaviour, and as such one can reasonably expect that the MOM scheme method would apply to these ensembles. However, the vortex-only results are dominated by infrared strength and a decay to $0$ at high momentum. Hence, a multiplicative renormalisation based on a perturbative expectation does not apply.

To approach this renormalisation issue, we present two sets of results. The first set will display all propagators from an ensemble divided by $Z_3^{\rm UT}$ as determined via the MOM scheme described in Eq.~(\ref{eq:MOMRenorm}). This allows us to readily compare the vortex-modified propagators across all ensembles.

Based on the findings of Ref.~\cite{Biddle:2018dtc}, we also consider renormalising the vortex-modified propagators via a best-fit approach. To do this, we consider taking a linear combination of the vortex-only and vortex-removed bare renormalisation functions, $Z^{\rm VO}_{\rm bare}(q^2)$ and $Z^{\rm VR}_{\rm bare}(q^2)$ respectively, such that the ``reconstructed'' propagator
\begin{equation}
    \label{eq:Recon}
  Z^{\rm recon}(q^{2}) = \frac{\zeta^{\rm VO}\,Z_{\rm bare}^{\rm VO}(q^2)+\zeta^{\rm VR}\,Z_{\rm bare}^{\rm VR}(q^2)}{Z_3^{\rm UT}}
\end{equation}
is fit to $Z^{\rm UT}(q^2)$ via a linear least-squares fit. Here, $\zeta^{\rm VO}$ and $\zeta^{\rm VR}$ are fit parameters defined such that the renormalised vortex-modified propagators are
\begin{align}
  Z^{\rm VO}(q^2)&=\frac{\zeta^{\rm VO}}{Z_3^{\rm UT}}\,Z_{\rm bare}^{\rm VO}(q^2) \label{eq:ZVO}\,,\\
  Z^{\rm VR}(q^2)&=\frac{\zeta^{\rm VR}}{Z_3^{\rm UT}}\,Z_{\rm bare}^{\rm VR}(q^2)\label{eq:ZVR}\,.
\end{align}
Fitting the reconstructed propagator is subject to the constraint
\begin{equation}
Z^{\rm recon}(\mu^2) = Z^{\rm UT}(\mu^2) = 1\,,
\end{equation}
so that the MOM scheme is replicated in the fit. This reduces the fit to a single parameter, as we can constrain e.g. $\zeta^{\rm VR}$ to be
\begin{equation}
  \zeta^{\rm VR}=\frac{Z_3^{\rm UT}-\zeta^{\rm VO}\,Z^{\rm VO}_{\rm bare}(\mu^2)}{Z^{\rm VR}_{\rm bare}(\mu^2)}\,.
\end{equation}
Once a fit is found, the renormalisation defined in Eqs.~(\ref{eq:ZVO}, \ref{eq:ZVR}) is applied such that the reconstructed propagator is simply given by the sum
\begin{equation}
  \label{eq:2}
  Z^{\rm recon}(q^{2})=Z^{\rm VO}(q^{2})+Z^{\rm VR}(q^{2})\,.
\end{equation}
As we shall see, this fitting approach is appealing as it produces excellent agreement between the pure-gauge untouched propagator and the reconstructed propagator as defined in Eq.~(\ref{eq:Recon}).

\subsection{Results}
\begin{figure}
  \centering
  \includegraphics[width=\linewidth]{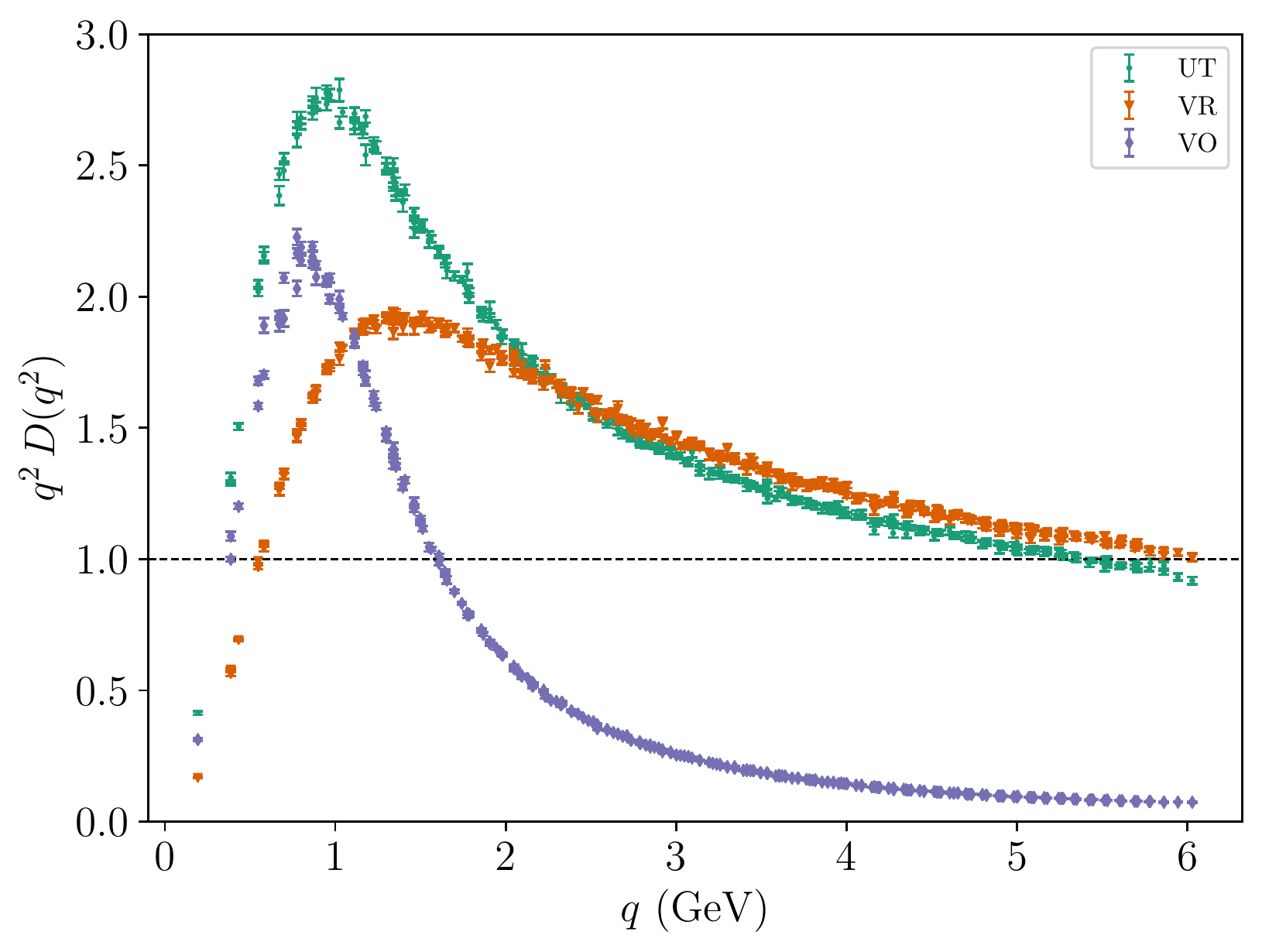}
  \caption{\label{fig:PGprop}Pure-gauge gluon propagator as calculated on the
    untouched (green), vortex-removed (orange) and vortex-only (purple)
    ensembles. All propagators a renormalised by applying the renormalisation
    constant found by applying the MOM scheme to the untouched propagator. A
    black line at $Z(q^{2})=1$ is used to show the asymptotic behaviour.}
\end{figure}

We first present the pure-gauge calculation of the scalar propagator, with all
results renormalised using the untouched renormalisation constant,
$Z_3^{\rm UT}$. The results from the three ensembles, UT, VO and VR are shown in
Fig.~\ref{fig:PGprop}. As expected, these results agree with those of
Ref.~\cite{Biddle:2018dtc}, with the untouched propagator defined by an infrared
peak and an ultraviolet plateau to tree-level. The vortex-modified counterparts
qualitatively capture these two features, with the vortex-only propagator
featuring an infrared peak, whereas the vortex-removed results retain the
ultraviolet plateau. However, there is still significant infrared strength
present in the vortex-removed propagator, which indicates that some long-range
physics remains in the vortex-removed ensemble.

\begin{figure}
  \centering
  \includegraphics[width=\linewidth]{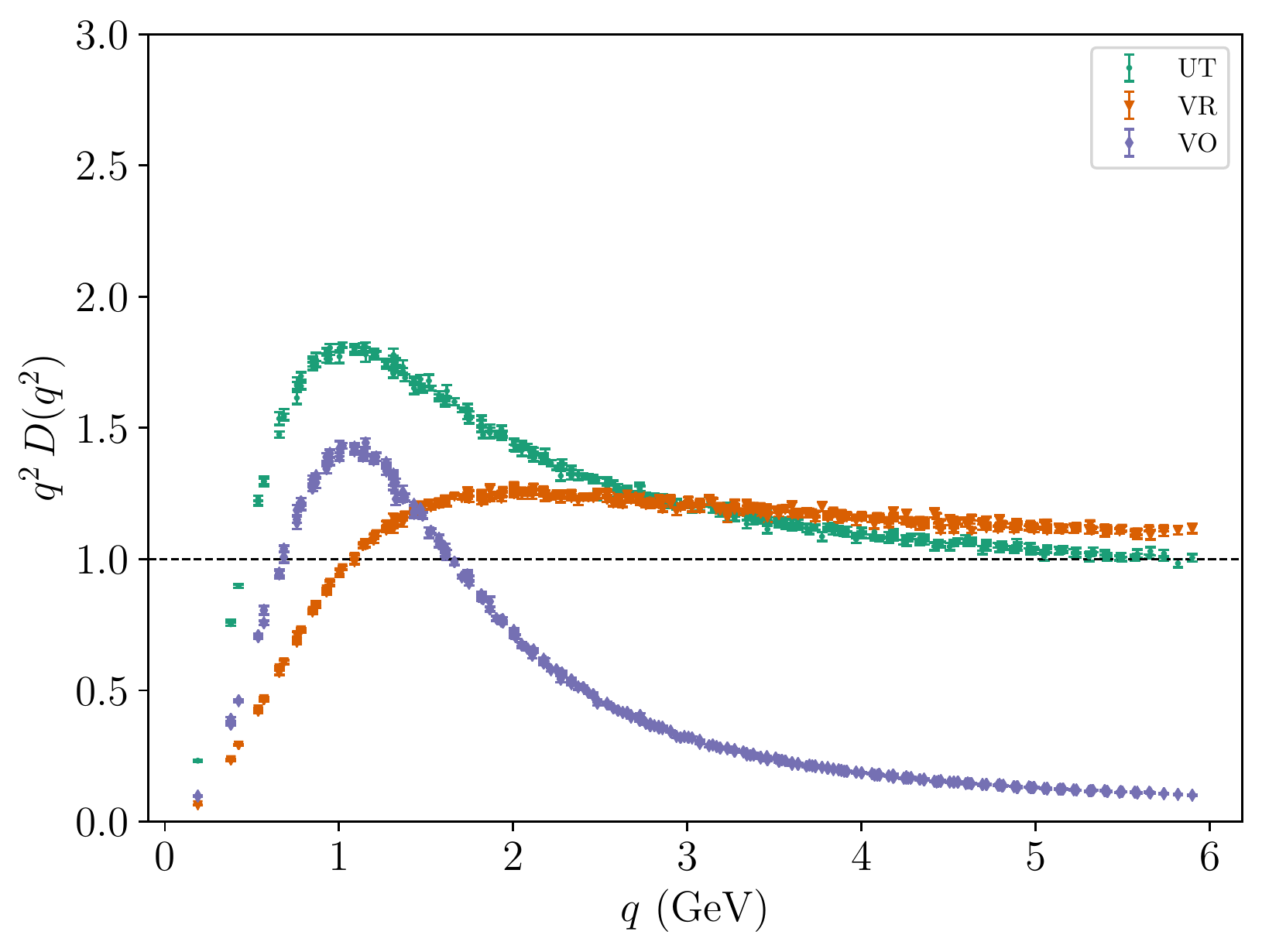}
  \caption{\label{fig:heavyprop}$m_{\pi}=701~\si{MeV}$ gluon propagator. The data scheme is
    as described in Fig.~\ref{fig:PGprop}.}
\end{figure}

We now consider the dynamical ensemble with the heaviest pion mass. We plot the gluon propagator calculated on this ensemble in Fig.~\ref{fig:heavyprop}. We observe that even at this unphysically large pion mass, the impact on the propagator is significant. Qualitatively, the propagators retain the same features as described for the pure-gauge sector, however the untouched propagator is noticeably screened, as is to be expected from the introduction of dynamical fermions~\cite{Bowman:2007du}. The vortex-only propagator also exhibits screening, which is a heretofore unseen effect. Furthermore, the infrared enhancement of the vortex-removed propagator is significantly reduced when compared to the pure-gauge results shown in Fig.~\ref{fig:PGprop}, and now displays behaviour completely consistent with the perturbative expectation. These two changes indicate a noticeable shift in the behaviour of the centre vortices under the introduction of dynamical fermions.

\begin{figure}
  \centering
  \includegraphics[width=\linewidth]{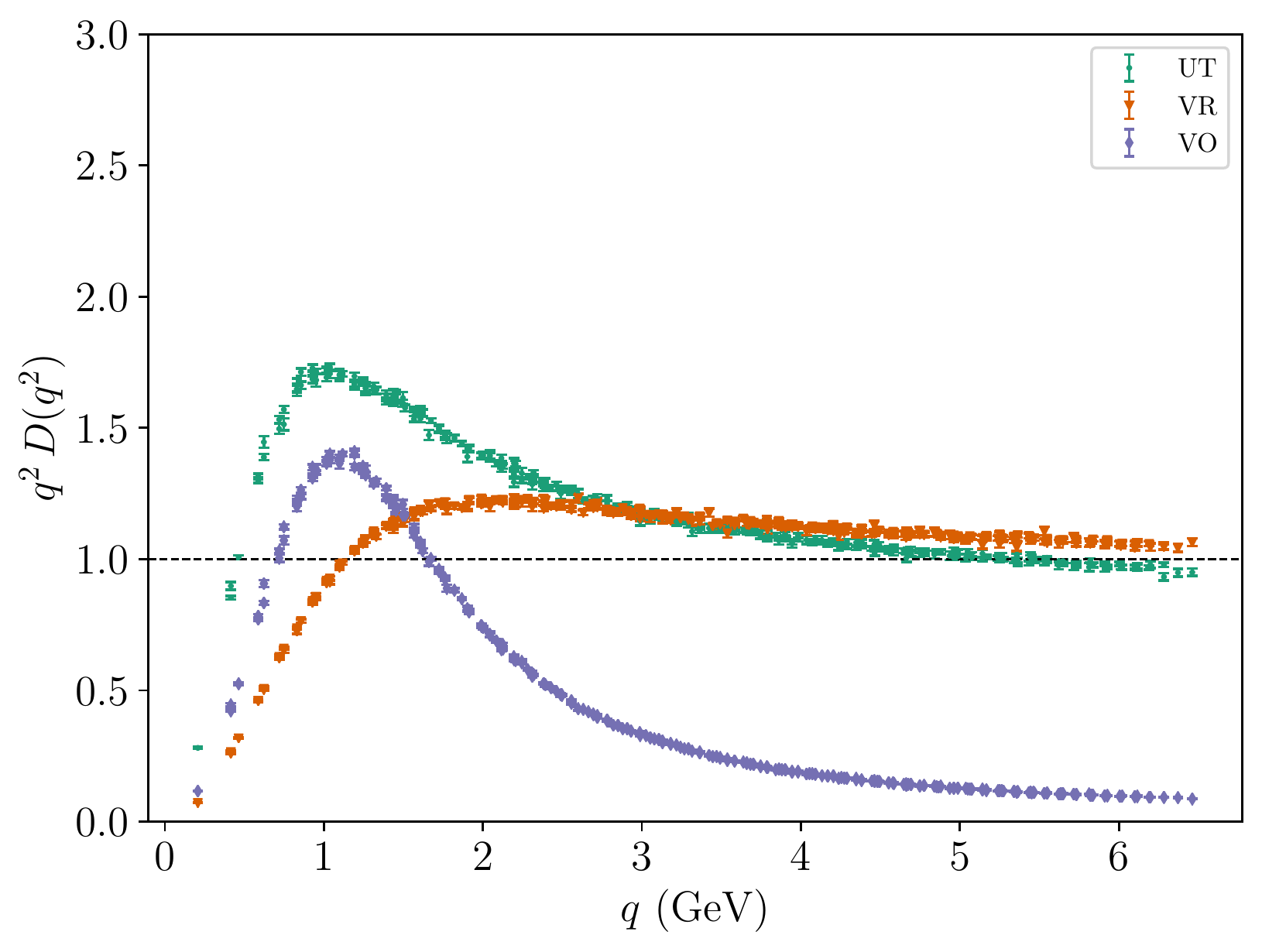}
  \caption{\label{fig:lightprop}$m_{\pi}=156~\si{MeV}$ gluon propagator. The data scheme is
    as described in Fig.~\ref{fig:PGprop}.}
\end{figure}

\begin{figure}
  \centering
  \includegraphics[width=\linewidth]{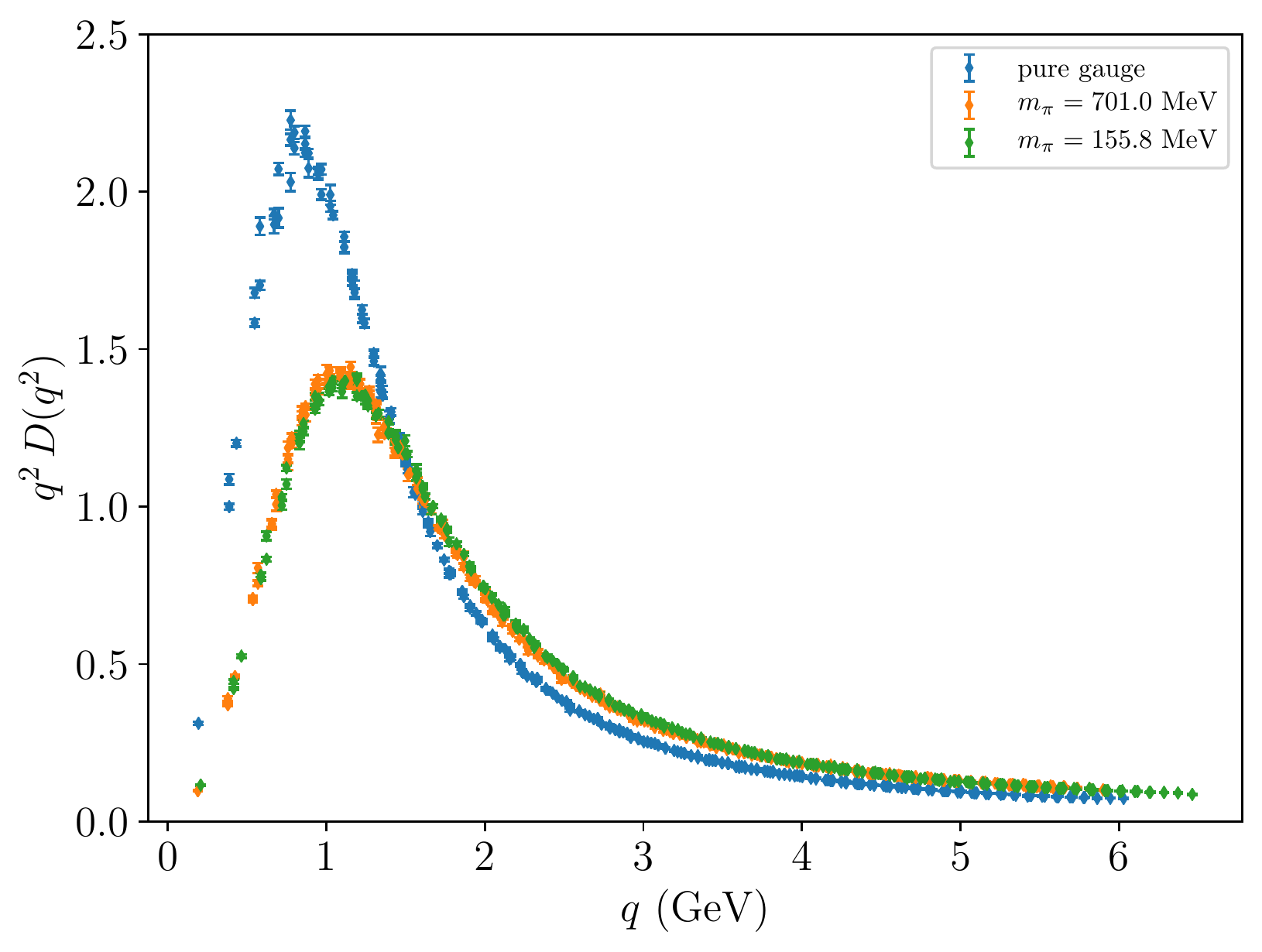}
  \caption{\label{fig:VOcomparison}The vortex-only propagators from all three ensembles. Screening is distinctly visible as we transition from pure-gauge to dynamical gauge fields.}
\end{figure}

The story is similar for the results of the lightest pion mass, shown in Fig.~\ref{fig:lightprop}. Screening effects are further enhanced in the untouched propagator as the pion mass is reduced, although it is difficult to observe any change in screening in the vortex-only propagator. To aid in this, we plot a comparison of the vortex-only propagators across all three ensembles in Fig.~\ref{fig:VOcomparison}. Here we can clearly see the presence of screening upon introduction of dynamical fermions. Between the two dynamical ensembles, screening effects are slightly enhanced as the pion mass decreases, however the effect is very subtle. The vortex-removed propagator also retains the suppression of infrared enhancement found at $m_{\pi}=701~\si{MeV}$. Given that the behaviour of the vortex-modified propagators is so similar between the two pion masses, it appears that the mere presence of dynamical fermions plays a substantial role in altering centre vortex structure and the manner in which they generate the gluon propagator.

An interesting trend in the results presented in Figs.~\ref{fig:PGprop}, \ref{fig:heavyprop} and \ref{fig:lightprop} is the fact that the vortex-removed results exceed the untouched results at high momentum with the same renormalisation constant applied. It is well understood that a larger renormalisation constant is necessary to account for increased roughness in an ensemble~\cite{Bowman:2002fe}. Given that the vortex removal process represents a significant change in the texture of the gauge field, it appears that such roughness has been induced in the vortex-removed fields. This finding supports the need for more detailed consideration of the renormalisation of the vortex-modified propagators.

\begin{figure}
  \centering
  \includegraphics[width=\linewidth]{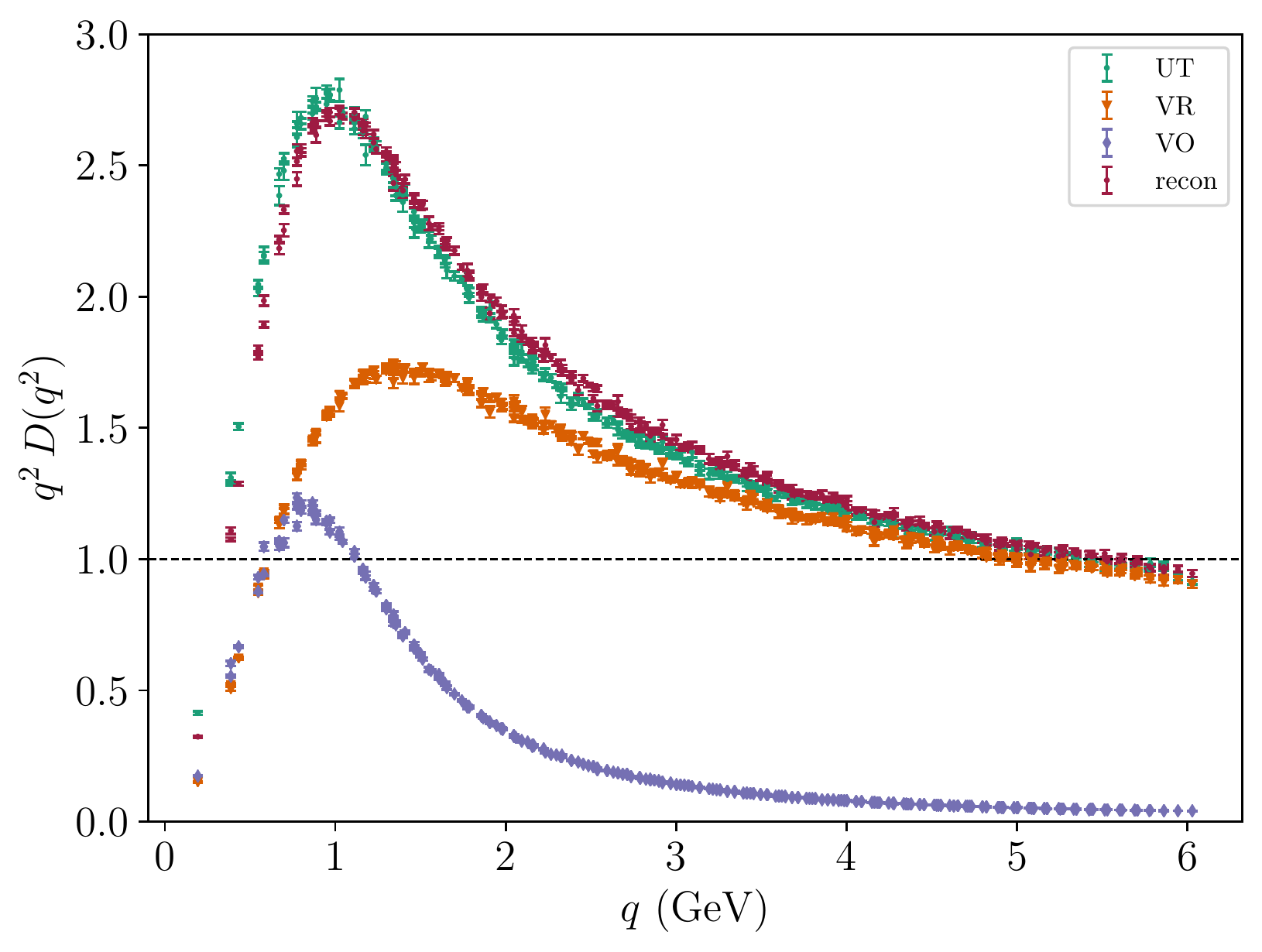}
  \caption{\label{fig:PGpropindep}Pure-gauge gluon propagator. All propagators are
    renormalised by applying the renormalisation method described at the end of
    Sec.~\ref{subsec:GlupropDefinition}. The ``recon'' data (red) is an
    attempted reconstruction of the original propagator by summing the
    vortex-only and vortex-removed propagators.}
\end{figure}

We now repeat the above presentation but with the second renormalisation method applied, as defined at the end of Sec.~\ref{subsec:GlupropDefinition}. The pure-gauge results are presented in Fig.~\ref{fig:PGpropindep}. The shape of the propagators is naturally the same as before, with the interesting addition from the renormalisation method being the reconstructed propagator. Here we observe good agreement between the untouched and reconstructed propagators. This indicates that the additional degree of freedom in the renormalisation method is to some extent encapsulating the manner in which the untouched propagator is partitioned into its vortex-modified components.

The dynamical ensembles with this renormalisation method applied show a reduced
agreement between the untouched and reconstructed propagators relative to the
pure gauge results. The significance of this disagreement is unknown, and
represents another interesting shift in behaviour when transitioning from
pure-gauge to dynamical QCD. The fit constants as described in
Eqns.~(\ref{eq:MOMRenorm}) and (\ref{eq:Recon}) are presented in
Table~\ref{tab:RenormConsts}.

\begin{figure}
  \centering
  \includegraphics[width=\linewidth]{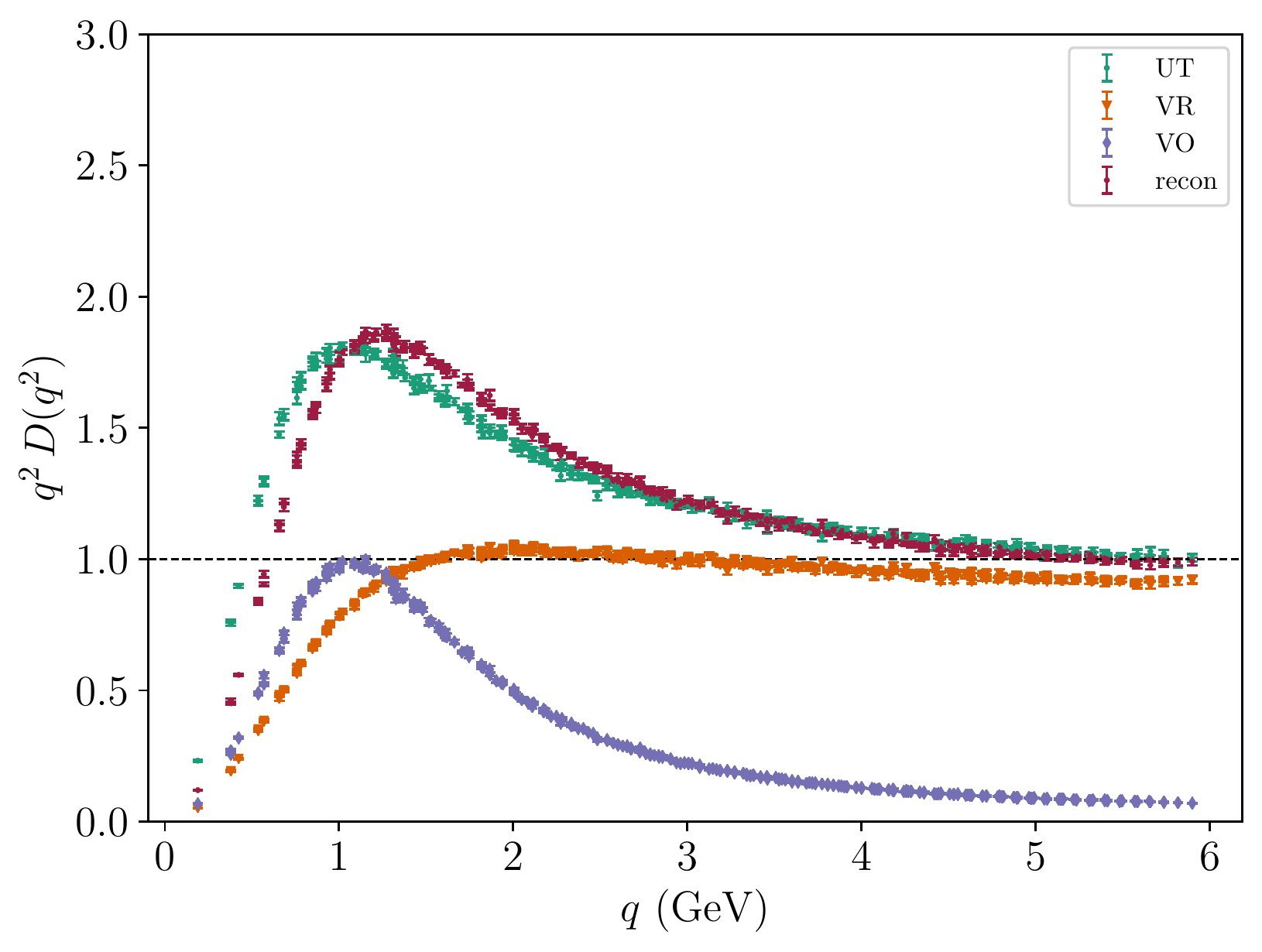}
  \caption{\label{fig:heavypropindep}$m_{\pi}=701~\si{MeV}$ gluon propagator. The data scheme is
    as described in Fig.~\ref{fig:PGpropindep}.}
\end{figure}

\begin{figure}
  \centering
  \includegraphics[width=\linewidth]{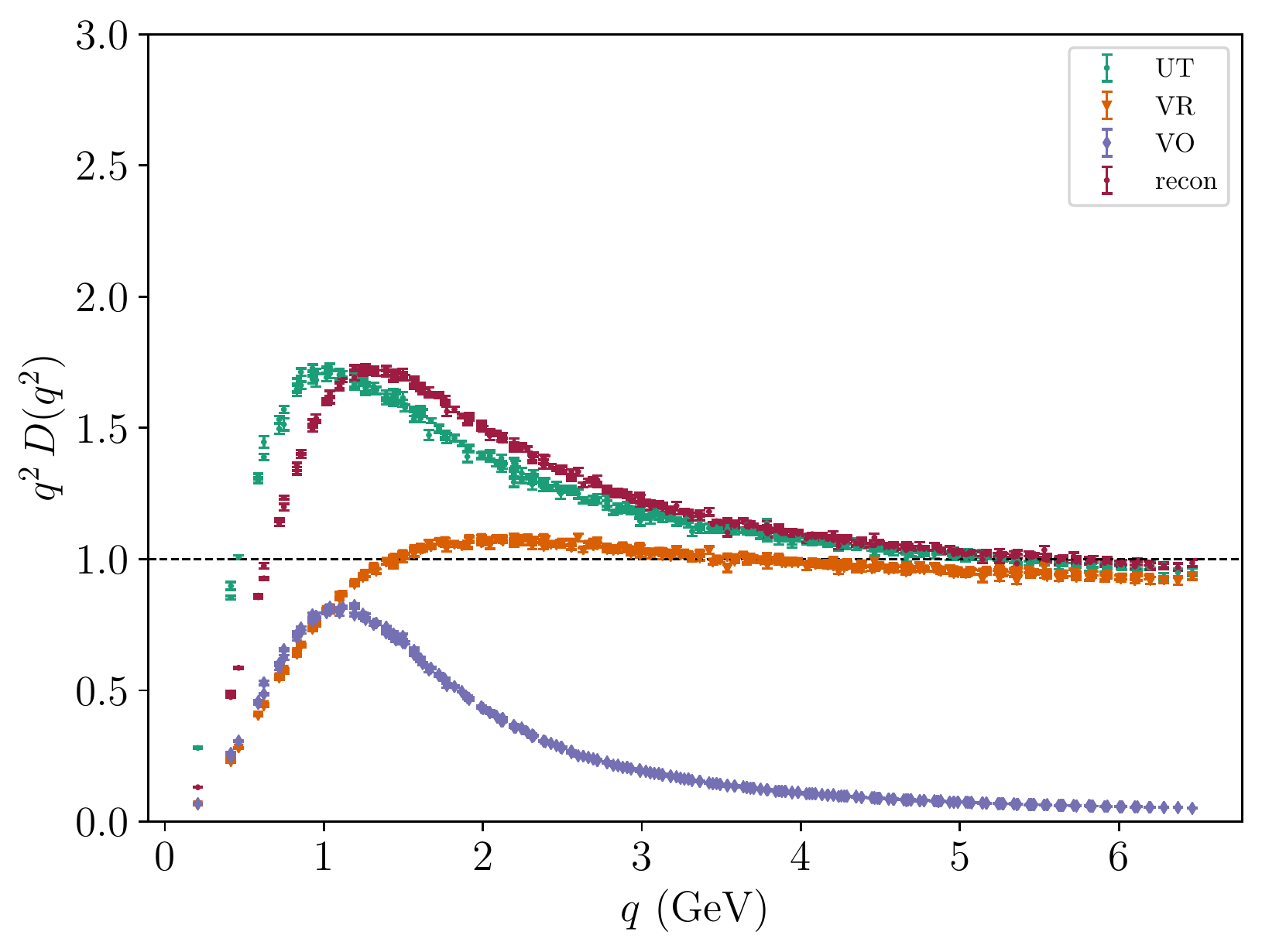}
  \caption{\label{fig:lightpropindep}$m_{\pi}=156~\si{MeV}$ gluon propagator. The data scheme is
    as described in Fig.~\ref{fig:PGpropindep}.}
\end{figure}

\begin{table}[tb]
  \caption{\label{tab:RenormConsts}The MOM scheme renormalisation constants,
    $Z_3^{\rm UT}$, as well as the fitted renormalisation constants defined in
    Eq.~(\ref{eq:Recon}).}
  \begin{ruledtabular}
    \begin{tabular}{cccc}
      Ensemble & $Z_{3}^{\rm UT}$ & $\zeta^{\rm VO}$ & $\zeta^{\rm VR}$ \\
      \hline\\
      Pure gauge & 7.112 & 0.5543 & 0.8985 \\
      $m_{\pi}=701~\si{MeV}$ & 9.316 & 0.6916 & 0.8251 \\
      $m_{\pi}=156~\si{MeV}$ & 11.51 & 0.5834 & 0.8780
    \end{tabular}
  \end{ruledtabular}
\end{table}

\begin{figure}[h!]
  \centering
  \includegraphics[width=\linewidth]{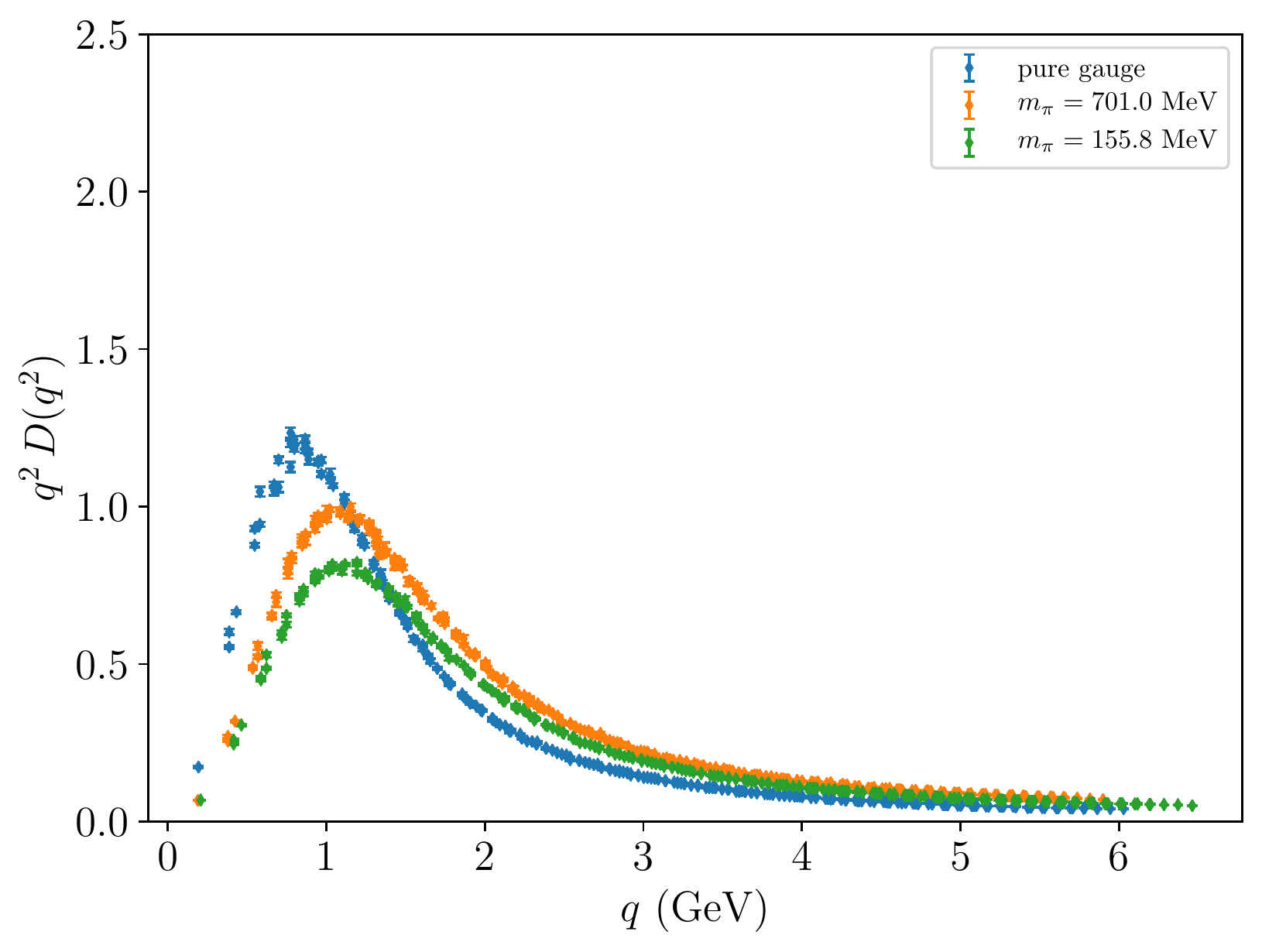}
  \caption{\label{fig:VOcomparisonindep}The vortex-only propagators from all
    three ensembles. The best fit renormalisation method produces a greater
    distinction between vortex-only propagators compared to
    Fig.~\ref{fig:VOcomparison}.}
\end{figure}

When comparing the vortex-only propagators with this new renormalisation scheme
we observe that screening effects remain apparent, as is evident from
Fig.~\ref{fig:VOcomparisonindep}. Furthermore, it is also possible to see a
distinct increase in screening behaviour as we transition from the heavy to
light pion mass. This suggests that perhaps this renormalisation method is more
representative of the relative contributions of the vortex-only and
vortex-removed propagators to the untouched propagator.

In summary, the vortex-modified propagators undergo significant changes in
behaviour upon the introduction of dynamical fermions. Residual infrared
strength present in the pure-gauge vortex-removed propagator is suppressed in
full QCD. The vortex-only propagators effectively capture screening effects
manifesting as suppressed infrared enhancement, indicating that the long-range
behaviour of the vortex-only fields mirrors their untouched counterparts.
Best-fit renormalisation provides further insight into the structure of these
vortex fields, where we find that the sum of vortex components reconstructs the
original propagator to a fair degree. This further supports the idea that the
vortex-only and vortex-removed propagators embody a splitting of the vacuum into
long- and short-range strength respectively.

\section{Positivity Violation}\label{sec:Positivity-Violation}
\subsection{Discussion}\label{sec:Positivity-Violation:Discussion}
For an arbitrary two-point function $D(x-y)$ to represent correlations between
physical particles in the sense of a Wightman quantum field
theory~\cite{Wightman:1956zz}, it is necessary by the Osterwalder-Schrader
axioms~\cite{Osterwalder:1973dx} for $D(x-y)$ to satisfy
\begin{equation}
  \label{eq:OS-Axiom}
  \int d^4x\,d^4y\,f^{*}(-x_0,\vect{x})\,D(x-y)\,f(y_0,\vect{y}) \ge 0\,,
\end{equation}
for a suitable complex test function $f$. If this axiom is satisfied, then the
scalar propagator defined in Eq.~(\ref{eq:Scalar-Propagator}) has spectral
representation
\begin{equation}
  \label{eq:kl-rep}
  D(p^2)=\int_0^{\infty}\,dm^2\,\frac{\rho(m^2)}{p^2+m^2}\,,
\end{equation}
with $\rho(m^2)\ge 0$, known as the K{\"a}llen-Lehmann representation.

To investigate the behaviour of the spectral representation, we consider the
Euclidean correlator, $C(t)$, obtained by taking the Fourier transform of
$D(p_0,\vect{0})$ as defined in Eq.~(\ref{eq:kl-rep}) such that,
\begin{equation}
  \label{eq:1}
  C(t)= \frac{1}{2\pi}\int_{-\infty}^{\infty}dp_0\int_0^{\infty}dm^2\frac{\rho(m^2)}{p_0^2+m^2}\,e^{-ip_0\,t}\, .
\end{equation}
Extending the $p_0$ integral to the complex plane and employing the residue
theorem, one arrives at
\begin{equation}
  \label{eq:3}
  C(t)=\int_0^{\infty}dm\, e^{-mt}\,\rho(m^2)\,.
\end{equation}
Clearly if $C(t)<0$ for any $t$ then $\rho(m^2)$ is not positive definite, and
we say that positivity has been violated. This implies that there is no
K{\"a}llen-Lehmann representation as defined in Eq.~(\ref{eq:kl-rep}), and as
such the propagator does not represent a correlation between physical states.
Hence, the states do not appear in the physical spectrum. In the context of the
gluon propagator, this can be taken as an indication that gluons are confined.

On the lattice~\cite{Leinweber:1998uu}, the Euclidean correlator, $C(t)$, is
given by the discrete Fourier transform of the temporal component of
Eq.~(\ref{eq:Scalar-Propagator}),
\begin{equation}
  \label{eq:Lattice-Correlator}
  C_{\text{lat}}(t)=\frac{1}{N_t}\sum_{n_t=0}^{N_t - 1} e^{-2\pi i n\,t/N_t }D(q_4(n_t)^2)\, ,
\end{equation}
where $N_t$ is the lattice extent in the temporal direction and $q_4$ is the
lattice momentum described in Eq.~(\ref{eq:Momentum-Substitution}) and
Eq.~(\ref{eq:Momentum-Variables}). As $D(0)$ is associated with the lowest
frequency mode of the propagator, it is a dominant term in
Eq.~(\ref{eq:Lattice-Correlator}). As such, it is essential to ensure that
finite volume effects are accounted for.

On the lattice, finite volume effects alter the tensor structure of the propagator given in Eq.~(\ref{eq:Gprop-Cont}) such that it has the general form~\cite{Leinweber:1998uu}
\begin{equation}
  \label{eq:Prop-Full}
  D_{\mu \nu}^{a b}(q)=\left(\delta_{\mu \nu}-\frac{h_{\mu \nu}(q)}{f\left(q^{2}\right)}\right)\, \delta^{a b}\, D\left(q^{2}\right)\, ,
\end{equation}
where $f(q^2)\rightarrow q^2$ and $h_{\mu\nu} \rightarrow q_\mu q_\nu$ for large $q_{\mu}$, but $f^{-1}(q^2)$ is finite at $q=0$. We define
\begin{equation}
  \label{eq:4}
  \tilde{h}_{\mu\nu}(q)=\frac{h_{\mu\nu}(q)}{f(q^2)}\,,
\end{equation}
and note that in the infinite volume limit,
\begin{align*}
  \label{eq:6}
  \tilde{h}_{\mu\mu}(q)&=f^{-1}(q^2)\,h_{\mu\mu}(q)\\
                       &=\frac{q_{\mu}q_{\mu}}{q^2}\\
                       &=\frac{q^2}{q^2}\rightarrow 1 \text{ as } q^2\rightarrow 0\,.
\end{align*}
However, on a finite volume lattice, $f^{-1}(q^2)$ cannot approach infinity. Since $q_{\mu}$ can take the value of $0$ and $f^{-1}(q^{2})|_{q^{2}=0}$ is finite, $\tilde{h}_{\mu\mu}=0$ for $q_{\mu}=0$ in a finite volume. Thus, the extraction of the scalar propagator $D(0)$ from the lattice propagator requires a normalisation different from that of Eq.~(\ref{eq:Scalar-Propagator}).

This change in normalisation can be implemented by noting that the quantity
\begin{equation}
  \label{eq:7}
  \left( \delta_{\mu\nu} - \frac{q_{\mu}q_{\nu}}{q^{2}} \right)\,,
\end{equation}
changes in the finite volume of the lattice to
\begin{equation}
  \label{eq:8}
  \left( \delta_{\mu\nu} - \tilde{h}_{\mu\nu}(q) \right)\,.
\end{equation}
For $q_{\mu} \ne 0$, setting $\mu=\nu$ and summing provides
\begin{equation}
  \label{eq:9}
  \sum_{\mu} \left( \delta_{\mu\mu} - \frac{q_{\mu}q_{\mu}}{q^{2}} \right)=4-1=3\,.
\end{equation}
But for $q_{\mu}=0$ on the lattice, $\tilde{h}_{\mu\nu}(0)=0$ and
\begin{equation}
  \label{eq:9}
  \sum_{\mu} \left( \delta_{\mu\mu} - \tilde{h}_{\mu\mu}(q) \right)=4\,.
\end{equation}
This results in
\begin{equation}
  \label{eq:0-Mom-Factor}
  D_{\mu\mu}^{aa}(0)=4\,(n_c^2-1)D(0)\,,
\end{equation}
as opposed to
\begin{equation}
  \label{eq:5}
  D_{\mu\mu}^{aa}(q)=3\,(n_c^2-1)D(q),~q\neq 0\,.
\end{equation}

To verify the validity of this factor we explore the behaviour of the ratio of off-diagonal to diagonal propagator components for $q_{\mu}=0$, i.e. ratios of the form
\begin{equation}
  \label{eq:Off-Diag-Ratio} \frac{D_{\mu\nu}(0)}{D_{\rho\rho}(0)}=\frac{\tilde{h}_{\mu\nu}(0)}{1-\tilde{h}_{\rho\rho}(0)},~\mu\neq\nu\,,
\end{equation}
where $\rho$ is not summed. As $\tilde{h}_{\mu\nu}\approx 0$, this ratio provides direct access to $\tilde{h}_{\mu\nu}$ relative to the Kronecker delta of $1$.

\begin{figure}
  \centering
  \includegraphics[width=\linewidth]{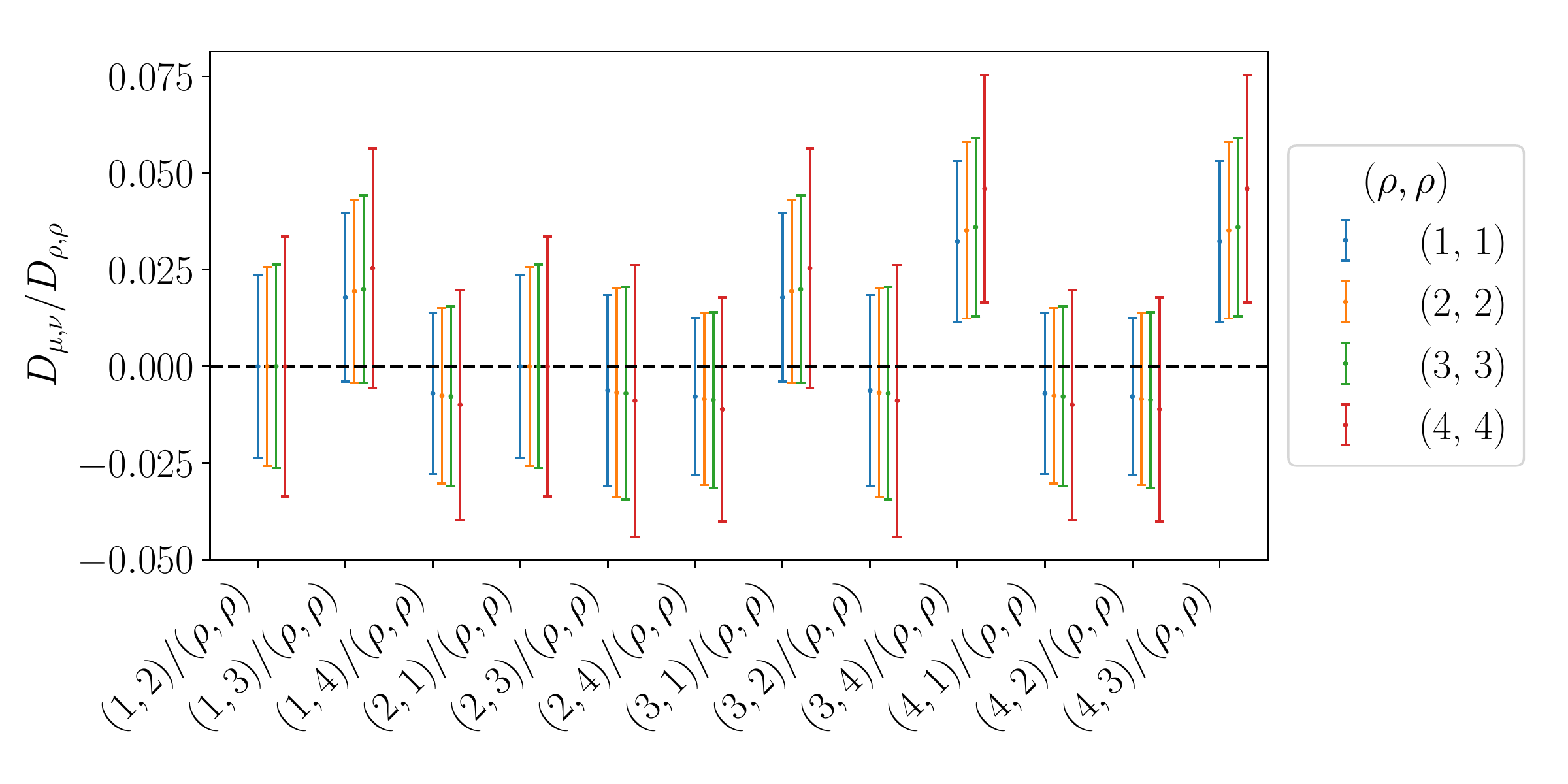}
  \caption{A plot of the 0-momentum ratio of the off-diagonal to diagonal tensor
    gluon propagator as described in Eq.~(\ref{eq:Off-Diag-Ratio}). We observe
    that the majority of values are consistent with zero, indicating that the
    lattice correction function $\tilde{h}_{\mu\nu}\rightarrow 0$ as
    $q\rightarrow 0$.}
  \label{fig:Gprop-Ratios}
\end{figure}

The values of these ratios calculated on the pure-gauge untouched configurations are shown in Fig.~\ref{fig:Gprop-Ratios}. It is clear that these ratios are consistent with 0 at $1\,\sigma$, indicating that both the diagonal and off-diagonal components of $\tilde{h}_{\mu\nu}$ are small relative to $1$. These results are corroborated by the other ensembles used in this work. This determination justifies the use of a factor of 4 instead of 3 in calculating the scalar propagator at zero momentum to address the impact of the finite volume.

\subsection{Results}
\begin{figure}
  \centering
  \includegraphics[width=\linewidth]{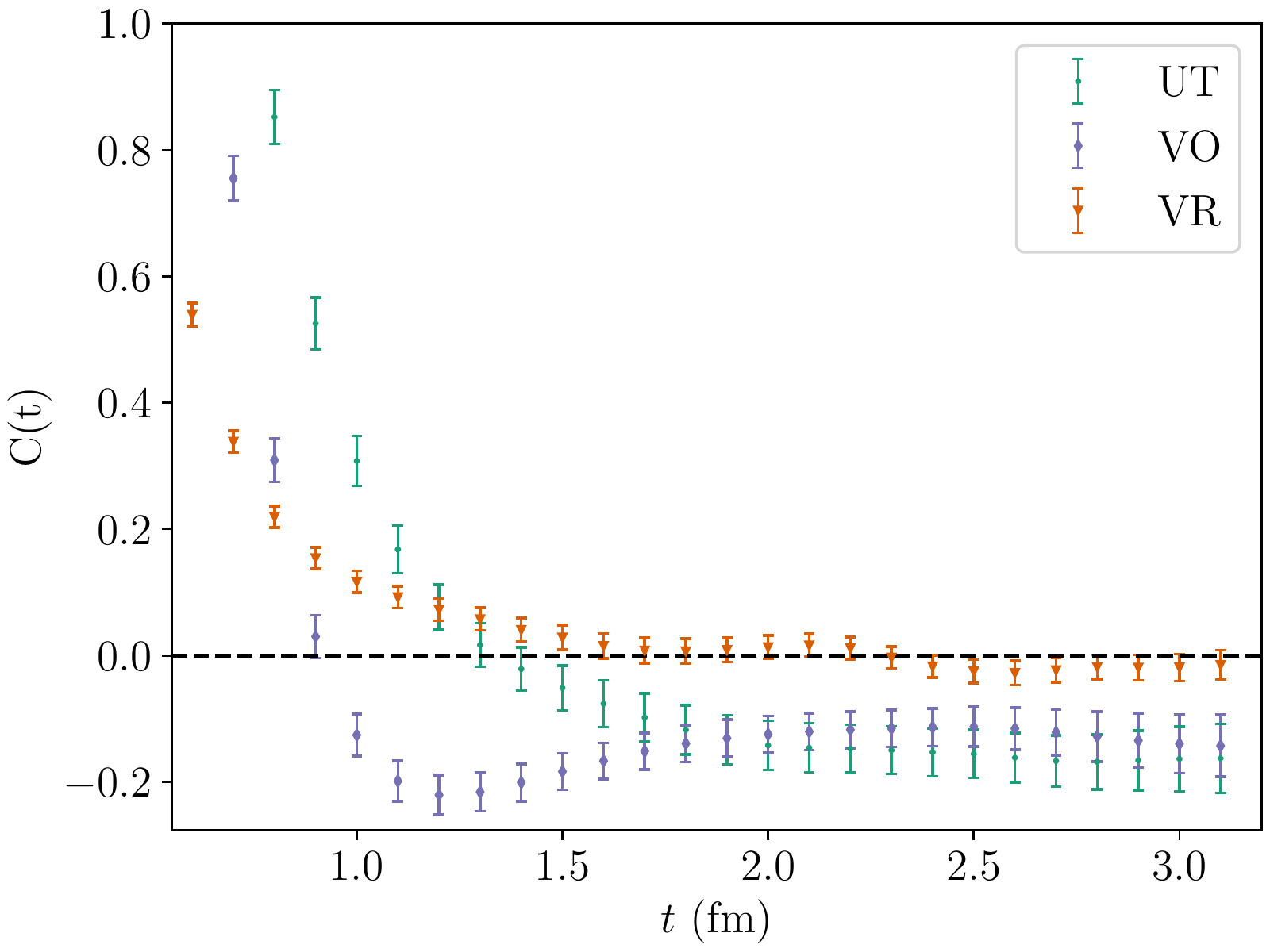}
  \caption{\label{fig:PGcorr}Pure-gauge Euclidean correlator. Shown are the
    results from the untouched (green), vortex-removed (orange) and vortex-only
    (purple) ensembles. A dashed line at $C(t)=0$ is provided to aid in
    observing positivity violation.}
\end{figure}

With this understanding developed, it is now possible to calculate $C(t)$ as defined in Eq.~(\ref{eq:Lattice-Correlator}). The results for the pure-gauge ensembles are shown in Fig.~\ref{fig:PGcorr}. As expected~\cite{Bowman:2007du}, the untouched correlator shows clear signs of positivity violation. Interestingly, the vortex-only correlators also exhibit robust positivity violation. The positivity violation present in the vortex-removed result at large distances is consistent with the observations made in Fig.~\ref{fig:PGprop}, where residual infrared strength in the vortex-removed gluon propagator is apparent. Thus, the separation of perturbative and non-perturbative physics through vortex modification is imperfect in the pure gauge sector.

\begin{figure}
  \centering
  \includegraphics[width=\linewidth]{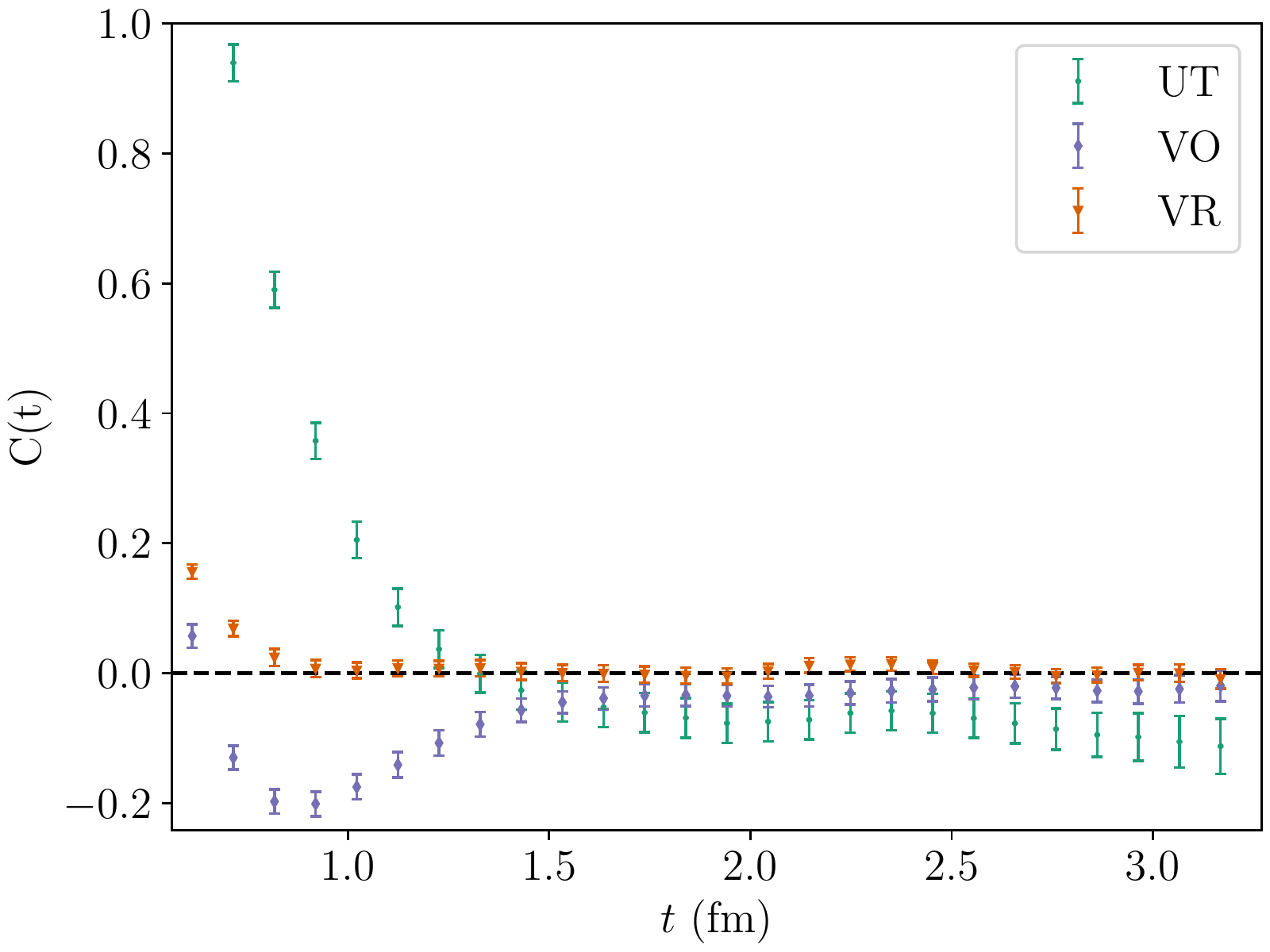}
  \caption{\label{fig:heavycorr}$m_{\pi}=701~\si{MeV}$ Euclidean correlator.
    Data is as described in Fig.~\ref{fig:PGcorr}.}
\end{figure}

\begin{figure}
  \centering
  \includegraphics[width=\linewidth]{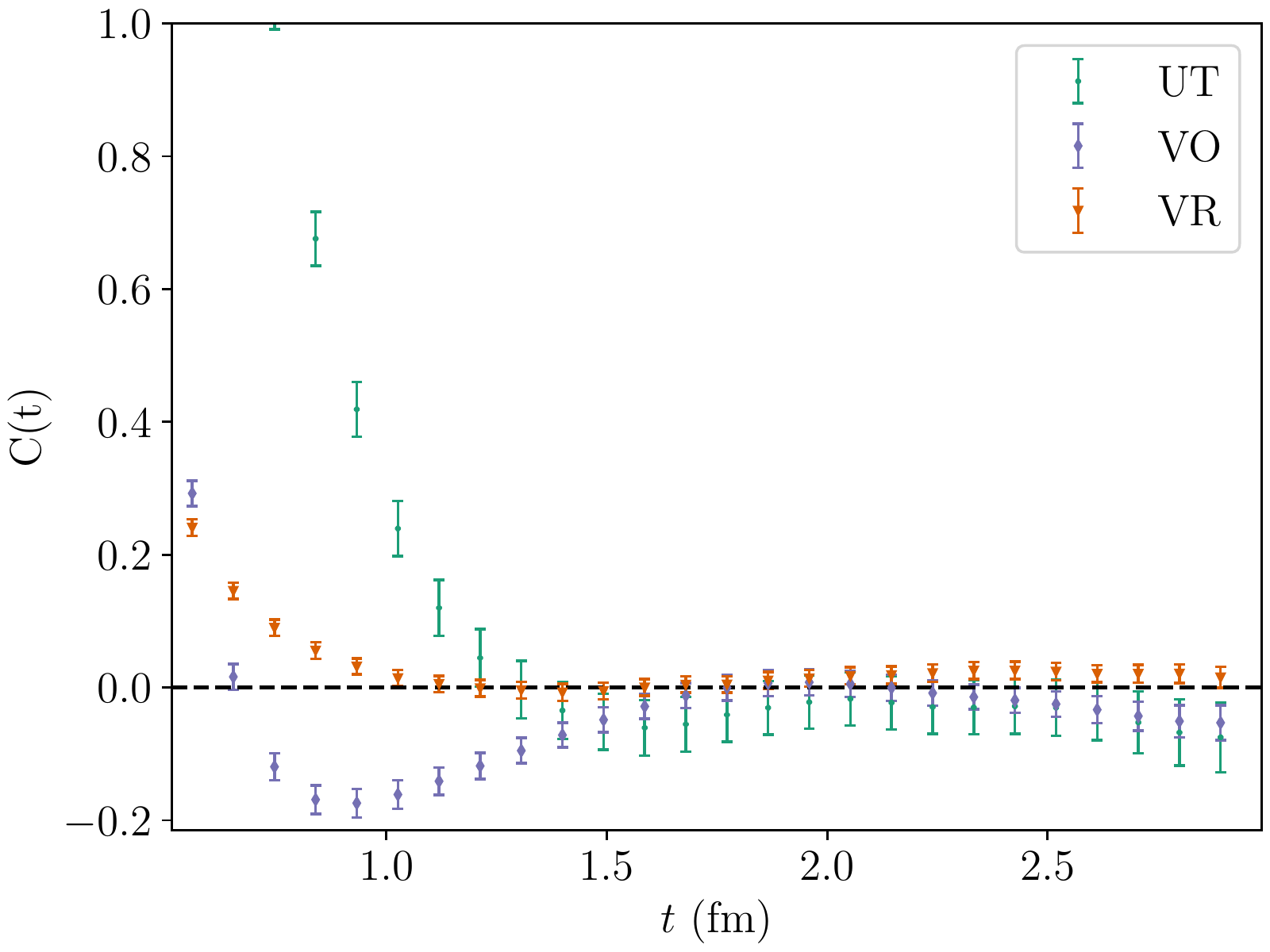}
  \caption{\label{fig:lightcorr}$m_{\pi}=156~\si{MeV}$ Euclidean correlator.
    Data is as described in Fig.~\ref{fig:PGcorr}.}
\end{figure}

The results from the dynamical ensembles, shown in Figs.~\ref{fig:heavycorr} and \ref{fig:lightcorr} demonstrate an interesting change in behaviour. Here we observe a similar robust violation of positivity in the vortex-only results as observed on the pure-gauge ensemble. However, the untouched results show a lesser degree of positivity violation, especially on the lightest pion mass ensemble shown in Fig.~\ref{fig:lightcorr}. Note however that violation is still present at large times.

As with the gluon propagator results in the previous section, the most striking change is in the vortex-removed correlator. In this sector we now observe consistency with positivity. This supports the interpretation of the positivity violation in the vortex-removed pure-gauge results as being related to the residual non-perturbative infrared strength in the gluon propagator. As this residual strength is significantly diminished on the dynamical ensembles, we now see that the residual $q^2$ dependence in the VR renormalisation function may be purely perturbative in origin. In this case, vortex modification has been successful in separating perturbative and non-perturbative physics.

In summary, vortex-only configurations exhibit significant positivity violation, as would be expected of a confining infrared-dominated theory. Conversely, the vortex removed configurations show a loss of this positivity violation, admitting the possibility that they do support a spectral representation of the propagator constructed from perturbative gluon interactions. These results provide additional support for the fact that centre vortices encapsulate the confining aspects of QCD.

\section{Conclusion}\label{sec:Conclusion}
Calculations of the behaviour of centre vortices in the presence of dynamical fermions are new, and each calculation provides further insight into the fascinating shift that centre vortices appear to undergo upon the introduction of dynamical fermions. Here we have found that centre vortices in the presence of dynamical fermions are effective in capturing the non-perturbative physics of QCD. Moreover, vortex removal appears to also be far more effective at removing the infrared strength of the propagator.

In regards to positivity violation, we establish the known result that unmodified lattice ensembles give rise to positivity violation in the Euclidean correlator~\cite{Bowman:2007du}. We then determined that both with and without the presence of dynamical fermions there is clear evidence that vortex-only ensembles exhibit significant positivity violation. On our pure-gauge ensemble, the vortex-removed correlator showed slight positivity violation at long distances, but on both dynamical ensembles this effect vanished. In full QCD, centre-vortex modification of the ground-state vacuum fields appears to provide an effective separation of perturbative and non-perturbative physics. These results present evidence that centre vortices in the QCD ground-state vacuum fields provide the origin of confinement.

The results presented here add to the growing number of investigations into the impact of dynamical fermions on centre vortices. We find marked differences between the pure-gauge and dynamical case, which implores further study. The origin of these disparities is currently unknown, and further work into the geometric structure of centre vortices in full QCD is planned. It is clear that there is an intimate relationship between dynamical fermions and centre vortices, and that this relationship has significant implications for the QCD vacuum.

\begin{acknowledgements}
  We thank the PACS-CS Collaboration for making their 2 +1 flavour
  configurations available via the International Lattice Data Grid (ILDG). This
  research was undertaken with the assistance of resources from the National
  Computational Infrastructure (NCI), provided through the National
  Computational Merit Allocation Scheme and supported by the Australian
  Government through Grant No. LE190100021 via the University of Adelaide
  Partner Share. This research is supported by Australian Research Council
  through Grants No. DP190102215 and DP210103706. WK is supported by the Pawsey
  Supercomputing Centre through the Pawsey Centre for Extreme Scale Readiness
  (PaCER) program.
\end{acknowledgements}

\bibliography{gluon_paper}
\end{document}